\documentclass[10pt]{iopart}

\usepackage{graphicx}
\usepackage{epsfig}
\usepackage{bm}

\def\be{\begin{equation}}
\def\ee{\end{equation}}
\def\bee{\begin{eqnarray}}
\def\eee{\end{eqnarray}}

\begin{document}



\title[Global gyrokinetic simulations of intrinsic rotation in ASDEX Upgrade Ohmic L-mode plasmas.]{Global gyrokinetic simulations of intrinsic rotation in ASDEX Upgrade Ohmic L-mode plasmas.}
\author{W.A.~Hornsby$^1$, C.~Angioni$^1$, Z.~X.~Lu$^1$, E.~Fable$^1$, I. Erofeev$^1$, R.~McDermott$^1$, A. Medvedeva$^1$, A. Lebschy$^1$, A.~G.~Peeters$^2$ and the ASDEX Upgrade Team}
\address{$^1$ Max-Planck-Institut f\" ur Plasmaphysik, Boltzmannstrasse 2, D-85748,Garching bei M\" unchen, Germany}
\address{$^2$ Theoretical Physics V, Dept. of Physics, Universitaet Bayreuth, Bayreuth, Germany, D-95447}

\ead{william.hornsby@ipp.mpg.de}

\date{\today}
\begin{abstract}

Non-linear, radially global, turbulence simulations of ASDEX Upgrade (AUG) plasmas are performed and the nonlinear generated intrinsic flow shows agreement with the intrinsic flow gradients measured in the core of Ohmic L-mode plasmas at nominal parameters.   Simulations utilising the kinetic electron model show hollow intrinsic flow profiles as seen in a predominant number of experiments performed at similar plasma parameters.  In addition, significantly larger flow gradients are seen than in a previous flux-tube analysis (Hornsby et al {\it Nucl. Fusion} (2017)).  Adiabatic electron model simulations can show a flow profile with opposing sign in the gradient with respect to a kinetic electron simulation, implying a reversal in the sign of the residual stress due to kinetic electrons.   The shaping of the intrinsic flow is strongly determined by the density gradient profile.   The sensitivity of the residual stress to variations in density profile curvature is calculated and seen to be significantly stronger than to neoclassical flows  (Hornsby et al {\it Nucl. Fusion} (2017)).  This variation is strong enough on its own to explain the large variations in the intrinsic flow gradients seen in some AUG experiments.  Analysis of the symmetry breaking properties of the turbulence shows that profile shearing is the dominant mechanism in producing a finite parallel wave-number, with turbulence gradient effects contributing a smaller portion of the parallel wave-vector.

\end{abstract}

\maketitle

\section{Introduction}
The quantitative prediction of intrinsic toroidal rotation caused by turbulent momentum transport  presents one of the major current challenges in the theoretical understanding of tokamak plasmas. It has been observed in multiple magnetically confined fusion experiments that, even without sources of external momentum, a confined plasma exhibits a toroidal rotation. Here the toroidal rotation profiles are determined largely by plasma transport processes.  This is known as intrinsic rotation and is of particular interest to ITER and to future reactor plasmas where the application of external torque will be comparatively small compared to the plasma inertia in contrast to existing tokamaks with high power neutral beam injection. 

Large intrinsic toroidal flow gradients have been measured in a variety of devices \cite{Rice07,camm16} and this, through the effect of flow shear, can have a stabilising effect on plasma turbulence \cite{Big90,Wal95,Stae13}, which causes the majority of heat and particle transport in magnetic fusion devices. Furthermore, high rotation and rotation shear is stabilising for large scale Magneto-hydrodynamic (MHD) modes \cite{Pol08,Rei07}.  Intrinsic rotation in L-mode shows complex behaviour.  One particularly interesting  observation is reversal of the intrinsic flow, from co-current to countercurrent and vice-versa, in the core with only very small variations in plasma parameters, in particular the plasma density \cite{bort06,rice08,rice11,ang11,reinke13,Camen16,Na16,shi17}.

Toroidal momentum transport is closely linked to symmetry breaking in the direction along the strong background magnetic field \cite{Pee05pop,Para11}.   At lowest order the gyro-kinetic equation is symmetric in the poloidal direction, $\theta$ and in the parallel velocity, $v_{||}$.  Momentum transport is intrinsically linked to mechanisms that break these symmetries.  

The total ion momentum flux, $\Pi_{i}$, can be written as a sum \cite {Pee11} of terms,
\begin{eqnarray}
\Pi_{i} = m_{i}n_{i}\big(V_{i,\phi}u 
+ \chi_{i,\phi}u' + M_{||}\gamma_{E}  + \Pi_{i,res}\big),
\label{momfluxeq}
\end{eqnarray}
where $u=R\Omega/v_{thi}$ is the ion Mach number and $u'=-(R^{2}/v_{thi})\nabla\Omega$, is the toroidal flow gradient.  $\Omega$ being the plasma toroidal angular rotation frequency and $v_{th i}$ is the ion thermal velocity, $v_{th i} = \sqrt{2 T_i / m_i}$.   $\chi_{\phi}$ is the toroidal viscosity and $V_{\phi}$ is the convective pinch velocity.  The diffusivities and pinch velocities are strong functions of the plasma turbulence, which in turn is a function of the density gradients, temperature gradients, magnetic equilibrium and collisionality.  The third term is the momentum flux generated by $E\times B$ shear flows, with shear rate, $\gamma_{E}$ \cite{waltz07,Barnes11}.  The residual stress component, $\Pi_{i,res}$, which is the momentum flux that is not related to the flow amplitude or its gradient, is particularly important as it is the source of intrinsic flow.  

A comprehensive and systematic validation of many residual stress mechanisms has been undertaken via extensive comparisons between a dedicated database of ASDEX Upgrade experiments and theoretical prediction.  The database comprises of $\sim{190}$ observations of Ohmic L-mode deuterium plasmas \cite{RMac14}.  In a recently published paper, quasilinear and non-linear local flux-tube simulations were used to study symmetry breaking mechanisms that are described in the local flux tube model \cite{Hor17}.  This included neoclassical background flow effects \cite{BarnesParra13}, up-down magnetic equilibrium asymmetry  \cite{Cam09pop,Cam09PRL,ball16}, higher order poloidal derivatives \cite{Sung13,Par15} (for example, the poloidal gradient of the electrostatic potential, $\partial\phi/\partial\theta$) and Coriolis effects \cite{Pee07prl,hahm07,PEEpop09,kluy09,weis12}.  While the sum of these symmetry breaking mechanisms predicted mostly hollow rotation profiles, as observed, they sustain gradients that are too small to describe completely the experimental profiles measured.  In the largest cases they achieved flow gradients reaching an amplitude of $u' = -0.4$ ($u' = -R_{0}/v_{th i}\partial u/\partial r$ where $R_{0}$ is the reference major radius).  However,  significantly larger gradients on the order of $u' = -1 \sim -1.5$ are regularly measured on AUG.  As such it is concluded that mechanisms which are not described by the flux-tube model could be responsible for this disparity.

There are a multitude of symmetry breaking mechanisms that require a global description of the turbulence including profile shearing \cite{camm11,Buchholz14}, the variation in the magnetic equilibrium \cite{LuNF15}, and turbulence intensity gradient effects \cite{gurcan10}.    This paper concerns itself with radially global simulations that investigate the effects of radial profiles on residual stress generation mechanisms utilising the experimental background profiles as input.  Global simulations of toroidal flow have been performed with similar gyro-kinetic codes, including simulations utilising GYSELA \cite{Sar11,Abit11}, XGC1 \cite{Ku12}, GYRO \cite{waltz11}, GTC \cite{holod1,holod2}  and recently with GTS \cite{wang10,Grierson17}.   Full-f gyro-kinetic simulations have also been performed with GT5D \cite{ido1,ido2} showing torque reversals linked to changes in the turbulence.

Unlike the previous study \cite{Hor17}, which applied a quasi-linear approach to the whole database, here we choose a small subset of data points to simulate due to the significantly larger computational costs of global simulations with respect to local flux-tube calculations.   Examples of the kinetic profiles and the intrinsic flow profiles are shown in Fig.~\ref{profiles}.  The intrinsic flow profiles seen in the database, and in general, fall into one of two categories.  Those that are hollow (blue curve in lower middle panel) with a strong flow gradient around mid radius, and those that are essentially flat, or slightly peaked (red curve).    It has been seen that within a single plasma discharge, both forms can be measured, and it has been suggested that a transition in the turbulence type is responsible for this flip \cite{angioni12,dia13}.  However this still remains an open question.

The rest of the paper is organised as follows, in Section \ref{model} the model and simulation set-up used are outlined.  In Sections \ref{sec151} a study on the global simulation of AUG discharge \#27000 and \#27001 is discussed and in Section \ref{d2n} the effects of the variation of the density profile on residual stress generation are discussed.  In Sec.~\ref{apptoexp} we discuss the application to experimental data and finally in Sec.~\ref{concs} we conclude.

\section{Model}
\label{model}

To study the effects of radial profiles on intrinsic flow generation  the global version of the gyro-kinetic code {\small GKW} is used.  Global here means that the radial variation of the magnetic equilibrium, the density and temperature profiles and the profiles in their gradients are considered. The gyro-kinetic set of equations \cite{brizard} is solved within {\small GKW}.   The full details of the code can be found in \cite{PEE09} and references found therein.  

The delta-$f$ approximation is used, where the background distribution function is assumed to be a Maxwellian ($F_M$), with particle density ($n$) and temperature ($T$).  $\rho_* = \rho_i / R_{0}$ is the normalised ion Larmor radius (where $\rho_i = m_i v_{th i} / e B_{0}$).  $R_{0}$ is the major radius,$B_{0}$ is the magnetic field on the magnetic axis and $a$ is the minor radius.  Neoclassical effects are neglected and there is no equilibrium flow in the simulations presented here. 

GKW uses straight field line Hamada \cite{HAM58} coordinates ($s,\zeta,\psi$) where $s$ is the coordinate along the magnetic field and $\zeta$ is the generalised toroidal angle. For circular concentric surfaces, the transformation of poloidal and toroidal angle to these coordinates is given by ($s,\zeta) = ((\theta + \epsilon \sin \theta)/ 2 \pi , [q \theta - \phi]/ 2 \pi)$.

\subsection{Krook operator}
\label{krook}

A Krook-like operator is used within the simulation domain to damp the perturbed distribution function so that a steady state is achieved.  The Krook operator has the form, 
\begin{eqnarray}
{\partial f (v_\parallel,\mu)\over \partial t} \biggr \vert_{K}  = - \gamma_K \biggl [   
{1 \over 2} [ f(v_\parallel,\mu) + f(-v_\parallel,\mu)] - \nonumber\\ \tilde n F_M(v_\parallel,\mu) \biggr ], 
\end{eqnarray}
where $\tilde{n}$ is the perturbed local density. $\gamma_{k}$ is a coefficient, chosen to be a fraction of the fastest growing mode growth rate, so as to minimise the effect on the turbulence while still maintaining the temperature profile close to its equilibrium form.  This operator conserves the local density and parallel momentum.  The same operator is used in layers near the boundaries to damp the distribution function to zero to minimise the influence of the boundary conditions on the solution.  

\subsection{Parameters and set-up.}
\label{parameters}

All simulations presented are set up in the following way.  Circular equilibrium geometry is used which is a good approximation in the core of ASDEX Upgrade.  The effect of magnetic equilibrium up-down asymmetry was performed in \cite{Hor17} and found to be small in this region.  All simulations are run with one ion species, in this case hydrogen, and kinetic electrons.  Impurities can have a stabilising effect on turbulence, however the boron impurity species is neglected  here for computational reasons.

\begin{figure}
\begin{centering}
\includegraphics[width=9.0cm,clip]{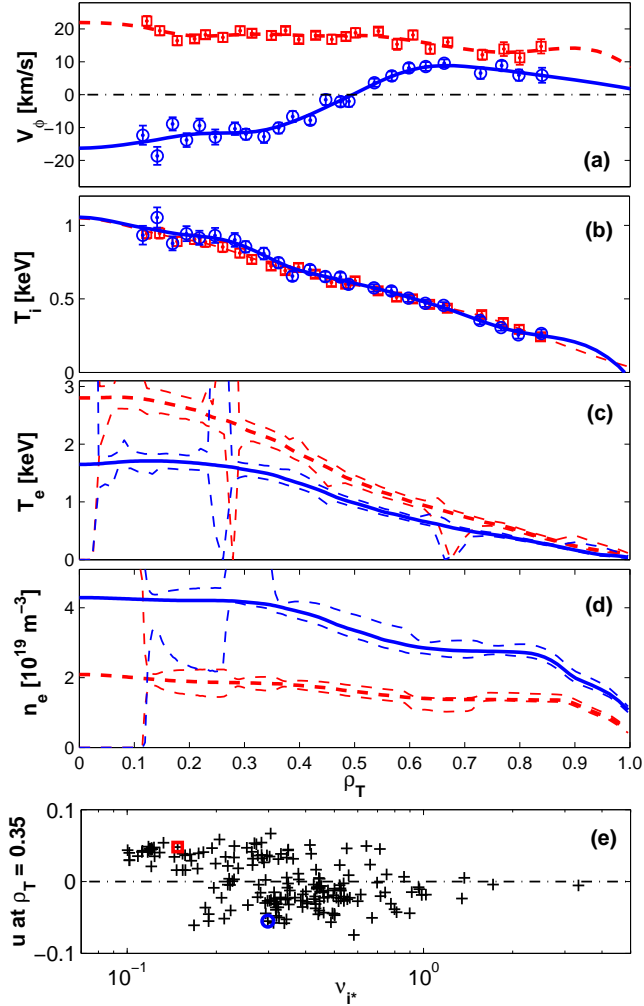}
\caption{The radial profiles of (a) toroidal flow profile, $u$, (b) ion and (c) electron temperature and (d) electron density for two discharges in the Ohmic database.  Red lines are from $t=2.1s$ from discharge number 27000 and blue lines
from $t=3.8s$ from the same discharge.  Dashed lines represent the confidence intervals in the profile reconstruction.  (e) The toroidal flow Mach number at $\rho_{T}=0.35$ for the
whole database, the two points plotted above are highlighted.  Adapted from Figure 1 of \cite{Hor17}. }
\label{profiles}
\end{centering}
\end{figure}

The number of radial grid points is chosen for each simulation to maintain enough resolution for the $\rho_{*}$ value chosen.  For example, a simulation with $\rho_{*}=0.005$ is run with $N_{x}=256$. The radial direction is treated using fourth-order finite-differencing with Dirichelet boundary conditions.      GKW uses a Fourier representation in the direction perpendicular to the magnetic field.  The toroidal wave vector is defined as
\begin{math}
k_\zeta \rho_i = 2 \pi n \rho_* n_{spac}.
\end{math}
$n_{spac}$ is the spacing between the toroidal modes, where $n_{spac}=1$ denotes a full torus simulation, however it is sometimes useful to use values larger than unity to maintain spectral resolution.  $n_{spac}=3$ for example would denote a simulation of a third of a torus.    The parameters used and the standard resolutions are summarised in Table 1.

\begin{table}
\begin{center}
\begin{tabular}{| l | l |}
\hline
  \multicolumn{2}{|c|}{Standard simulation parameters.} \\
\hline
Mass ratio, $m_{i}/m_{e}$ & 1836 \\
Num. parallel grid pts., $N_{s}$ & 16 \\
Num. parallel velocity grid pts., $N_{v_{||}}$ & 64 \\
Num. magnetic moment grid pts., $N_{\mu}$ & 16 \\
Max value $v_{||}$ ($v_{th i}$) & 4 \\
Max value $\mu$ ($v_{th i}^{2}/2B_{0}$) & 8 \\
&\\
{\bf Radial domain} & \\
$\psi_{min}=\frac{r_{min}}{R_{0}}$, ($\frac{r_{min}}{a}$) & 0.01, (0.03) \\
$\psi_{max}=\frac{r_{max}}{R_{0}}$, ($\frac{r_{max}}{a}$) & 0.245, (0.8) \\
&
\\
\hline
\end{tabular}
\label{simparams}
\caption{Table of standard parameters used in all simulations shown in this paper.}
\end{center}
\end{table}

The simulations are run without any equilibrium rotation and are initialised with low level noise with zero toroidal flow.  It was previously shown that GKW conserves the canonical angular momentum \cite{Buchholz14}.  Here we neglect the parallel velocity nonlinearity and higher order parallel derivatives \cite{Sung13} which have been shown to produce negligible parallel momentum at values of $\rho_{*}$ seen in AUG L-mode plasmas.  The pitch-angle electron-ion scattering operator is used.  The collision frequency  is set in our simulations by the experimental value of the density, temperature and major radius at mid radius in the tokamak.  The electron-ion collision frequency is defined as:

\begin{math}
\nu_{ei} = \frac{Z_{e}^{2}Z_{i'}^{2}e^{4}\ln{\Lambda_{ei}}n_{i}}{4\pi \epsilon_{0}^{2}m_{i}^{1/2}T_{e}^{3/2}},
\end{math}
where  $\ln{\Lambda_{ei}}$ is the Coulomb logarithm for scattering species $i$ and scattered species, $e$, $Z_i$  is the relative charge.  

The flux-surface averaged, gyro-centre fluxes of the toroidal momentum and heat shown in this paper are defined as:
\begin{eqnarray}
\Pi_{i} = \Pi_i^\psi  = \biggl \{ \int {\rm d}^3 {\bf v} \frac{s_{B}RB_t}{B}mv_\parallel\tilde {\bf v}_{E}\cdot \nabla \psi   f \biggr \} \\
Q_{i} = Q_i^\psi  = \biggl \{ \int {\rm d}^3 {\bf v}\frac{v^2}{2} \tilde {\bf v}_E\cdot \nabla \psi   f \biggr \}.
\end{eqnarray}
$\nabla\psi$ is the gradient in the normalised radial coordinate, $\psi=r/R_0$ and ${\bf v}_{E}$ is the $E\times B$ velocity .  Curly brackets denote a flux surface average.

\section{Global calculation of intrinsic flow with realistic gyro-radius}
\label{sec151}

A large number of symmetry breaking mechanisms are reliant on the radial variation of equilibrium kinetic and geometric quantities.  These include profile shearing which is linked to the tilting of mode structure by the radial variation of equilibrium \cite{camm11}.  These can not be captured in a flux tube, local simulation and hence require global gyro-kinetic simulations to evaluate their amplitudes. Due to the increased computational costs of these simulations we concentrate on a reduced dataset corresponding to a ASDEX Upgrade density ramp experiment which was performed over two consecutive discharges, \#27000-27001. The discharges were operated in the L-mode, with magnetic field strength on axis of $B = 2.5$ T and a plasma current of $I_{p} = 1.05$ MA.  An example of the profiles used in the simulation and an overview of the parameters seen in the ASDEX Ohmic L-mode rotation database is seen in Fig.~\ref{profiles}.  The flow profiles (a) and the ion temperature profiles (b) are measured by charge exchange recombination spectroscopy (CXRS)  \cite{macCXRS}.  The electron temperature profiles (c) are provided by electron cyclotron emission radiometry, and the local measurements are fitted within the integrated data analysis (IDA) applied at AUG \cite{fisIDA}. Also the electron density profiles (d) used in this study are obtained from IDA and are based on the combination of line integrated interferometry measurements in the core and local edge measurements from a lithium beam diagnostic in the edge.  As such the electron density profile shapes can have large uncertainties. The dashed lines in Fig. 1 (c,d) show the confidence intervals of the IDA reconstructed profiles.

\begin{figure*}[ht]
\begin{center}
\includegraphics[width=16.5cm,clip]{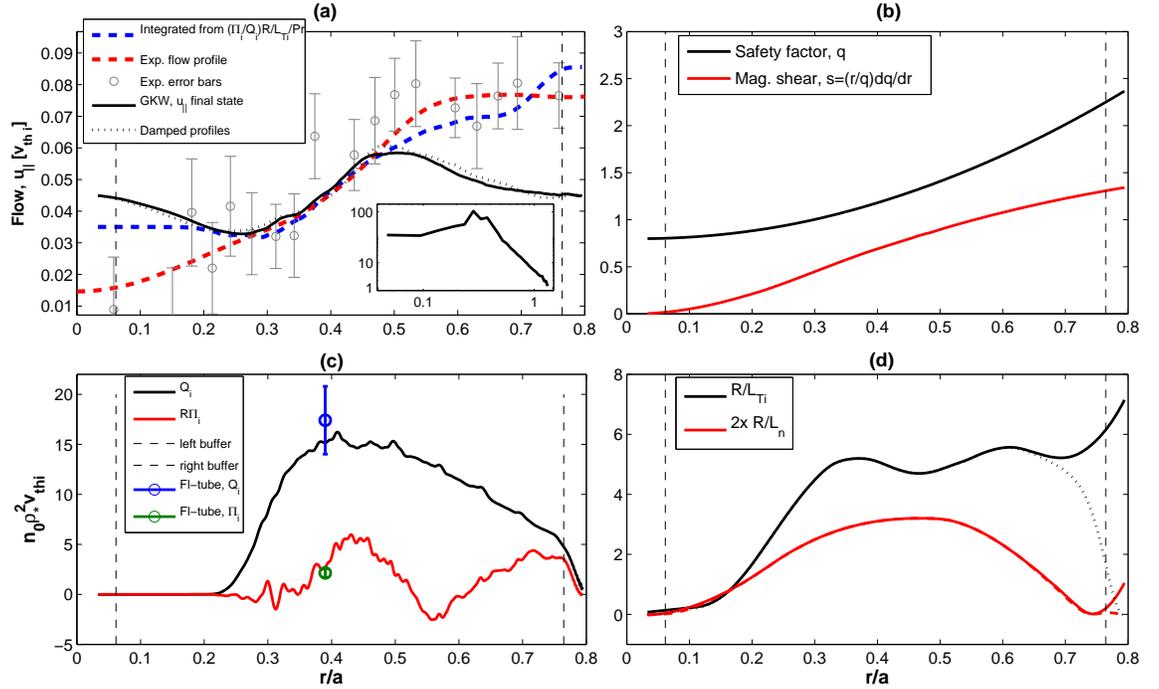}
\caption{(a) The radial profiles of the turbulence generated flow profile. In inlay is the time averaged and spatially averaged turbulent spectrum ($|\phi|^{2}$) of the electrostatic fluctuations as a function of the toroidal mode wave vector ($k_{\zeta}\rho_{i}$).  (b) The radial profiles of the safety factor (black) and magnetic shear ($\hat{s}=(r/q)\partial q/\partial r$, red). (c) The time averaged radial heat (black) and momentum  flux (red) profiles just after the linear overshoot of the simulation.  The vertical dashed lines denote the buffer regions where the turbulence is damped away near the boundaries. Red and green points are the heat and momentum fluxes from a flux-tube simulation at that radial position.  The error bars represent the standard deviation in the flux fluctuations. (d) The radial profiles of the ion temperature (black) and density (red) scale lengths for discharge number \#27001, $t=2.8s$.  Dashed profiles are those modified to reduce turbulence at the boundaries in a separate simulation.}
\label{flows}
\end{center}
\end{figure*}

In the simulations, the final flow state is determined by the achievement of a stationary state.  This occurs when there is no net momentum flux through any specific flux surface ($\Pi_{i}=0$) \cite{parra12}.  In these simulations this is achieved when the diffusive momentum flux caused by the non-linearly generated flow gradients balances, point-by-point radially, the residual stress component (in these global simulations the Coriolis effects are neglected to isolate the effects of the residual stress component.).   To reach this state the code needs to be run for a minimum of $t\sim 2000 R_{0}/v_{thi}$ time units which is significantly longer than the collision time.

Gradient driven simulations are run with kinetic electrons over a wave-number range encapsulating the ion-scale turbulent modes.  The equilibrium background ion-temperature and density and safety factor profiles are the experimentally measured ones.  Multi-scale simulations that include short wavelength ETG modes would increase the computational costs dramatically and are, therefore, not considered here.  However, it has been shown that ion flows are predominantly excited by turbulent fluctuations that have small wave-numbers ($k_{\zeta}\rho_{i}\leq 1.0$) \cite{tsd17}.  As such, for this study, higher wave-numbers need not be resolved.  The simulations are run electro-magnetically with a small value of the electron $\beta$ ($3\cdot10^{-4}$) to maintain numerical stability while having a negligible effect on the transport coefficients.  The radial grid spacing is chosen so that three grid points per gyro-radius resolution ($\Delta r/\rho_{i} = \Delta\psi/(N_{x}\rho_{*}) \sim 0.31$) is maintained.   The gyro-average is performed over a 64 point ring.   An example of the spectrum of electrostatic fluctuations is shown in inlay in panel (a) of Fig.~\ref{flows}.  The parameters and resolutions used are outlined in Table 2.

\begin{table}
\begin{center}
\begin{tabular}{| l | l |}
\hline
  \multicolumn{2}{|c|}{Simulation parameters.} \\
\hline
Discharge \# & 27001 \\
Time & 2.8s\\
Magnetic field on axis, B & 2.5T\\
Plasma current, $I_{P}$ & 1.05MA\\
Normalised gyro-radius, $\rho_{*}$ & 0.0015 \\
Number of radial grid points, $N_{x}$ & 512 \\
Number of toroidal modes, $N_{\zeta}$ & 29 \\
Mode spacing, $n_{spac}$ & 5 \\
Smallest wave-number, $k_{\zeta}\rho_{i}$  & 0.047 \\
Largest wave-number,  $k_{\zeta}\rho_{i}$  & 1.32 \\
$T_{e}/T_{i}$ at $r/R_{0}=0.12$ & 1.3\\
Ref. density, $n_{e}$ at $r/R_{0}=0.12$ & 5.05 ($10^{19}m^{-3}$)\\
Ref. temperature, $T_{i}$ at $r/R_{0}=0.12$ & 0.550 keV \\
Major radius, $R_{0}$ & 1.67m \\
&\\
\hline
\end{tabular}
\label{simparams2}
\end{center}
\caption{Table of parameters used in simulation with $\rho_{*}=0.0015$.}
\end{table}

Firstly, we concentrate on a time slice late in discharge number \#27001,   which is a high density time point in the density ramp.  Here the predominant linear mode at $k_{\zeta}\rho_{i}=0.42$ is the ITG mode.  The growth rate and mode frequency is calculated to be $\gamma = 0.145 v_{thi}/R_{0}$ and $\omega = 0.76 v_{thi}/R_{0}$ at $\psi = 0.13$.    The measured flow at this time point is co-current with a sharply hollow profile which has a maximum $u' = -0.55$ at $r/a = 0.45$.  This is shown in panel (a) of Fig~\ref{flows}.  The radial ion temperature gradient, density gradient, safety factor (q) and the magnetic shear ($\hat{s}$) profiles for this time-slice respectively, are shown in top right (b) and bottom right (d) panels of Fig.~\ref{flows} respectively.   

Panel (a) of Fig.~\ref{flows} shows the radial profiles of the parallel flow (the first parallel velocity moment of the perturbed distribution function) as calculated by GKW in black.  It shows a dipole like structure which is enforced by the no-slip boundary conditions and the conservation of momentum.  Also plotted is the fitted experimental intrinsic flow profile (red dashed line) and the corresponding raw data points (in light grey) and error bars.    It should be pointed out here that in the experiment, it is the boron rotation profile which is measured, but here modelled is the main ion rotation. However the difference is known to be small, particularly in the intrinsic flow gradient.  Moreover, it has been shown that the impurity momentum flux can be significant \cite{stae14}, however here it is small due to the low concentration of boron (around $\sim 2\%$ at the lowest densities and decreasing with increasing density) in these discharges.

\subsubsection*{Boundary condition effects.}

It is well known, and was shown in the previous section, that the boundary conditions in global simulations can have a dramatic effect on intrinsic flow generation \cite{Abit11,Ku12,Syi}.  As such, care should be taken when interpreting a flow profile from simulations and comparing to experiment.  The flow profile is heavily restricted by the boundary conditions at the edge of the computational domain, which set the flow to zero.  Furthermore a damping region is used at the edge of the computational domain to minimise the effects of the boundary conditions and these areas are denoted by vertical dashed lines.  Due to these combined effects one can not directly compare the flow and its gradient to experiment across the whole simulated domain.

As a test of whether the boundaries have a dominant effect on the flow shape and amplitude, the kinetic gradients are multiplied by a damping function so that the gradients tend to zero smoothly towards the boundary (these are plotted as dashed lines in panel (d) of Fig.~\ref{flows}).  This has a negligible effect on the gradient profiles.  At the inner boundary, the boundary conditions effect is naturally zero as there is no turbulence.  On the outer boundary there is marked reduction in the turbulence amplitude due to the reduction in the drive on the linear mode, however this is localised to the buffer layer.   It is seen that this has a very small effect on the flow profile, which is plotted as a dashed black line in panel (a) of Fig.~\ref{flows}.  In the centre of the domain, the effect is almost negligible implying that the torque from the boundaries has a very small effect on the flow gradient in the middle of the domain.  The invariance could be related to the relatively unchanged density profile which has a large effect on the residual stress.  This will be further investigated later in this paper.

\subsubsection*{Comparison.}

It is seen that the flow, and its gradient, are free in the centre of the domain to attain the value as determined by the turbulence. It can be seen here that the simulated flow profile and gradient agree extremely well with the experiment.  It should be noted that in the previous study using flux-tube simulations the sum of all the gradients at mid-radius was very small ($u'=-0.043$) for this case \cite{Hor17}.  As such global mechanisms would be expected to make up most of the residual stress, as is seen here. 

It is also possible to calculate the intrinsic flow in a complementary way, assuming that the total toroidal momentum flux is made up of only the diffusive (ion viscous) component and a residual stress, and that these balance each other to produce a flow gradient in the (normalised) form,\begin{equation}
u'(r) = -\frac{v_{th i}}{R_{0}}\frac{\Pi_{i,res}}{2Q_i}\frac{R/L_{Ti}}{Pr}.
\end{equation}
Here, all the quantities are functions of minor radius.   The radial heat flux is written in the local form, $Q_{i} = -n_{i}\chi_{i}\nabla T$ where $\chi_{i}$ is the turbulent heat diffusivity. 
The Prandtl number is defined as the ratio of the momentum and heat diffusivities, $P_{r} = \chi_{\phi,i}/\chi_{i}$.   The residual flow profile can be calculated from early time momentum flux, isolating the residual stress component just after the linear overshoot and before the zonal and toroidal flows are established in the simulation which can have a large effect on the momentum flux.  This method has a drawback in that the fully turbulent state might not have been reached at this time.  The residual stress profile can then  be integrated to produce a flow profile which is plotted as a blue dashed line in Fig.~\ref{flows}.  The procedure was followed in \cite{Grierson17} to calculate and compare the flow with DIII-D experiments.

This flow profile shows excellent agreement with the experimental flow profile, being largely within the error bars of the measurement and also agrees with the flow profile as calculated in the final flow state in Fig.~\ref{flows} (panel (a)), showing internal consistency.  This flow is calculated from turbulent quantities (heat and momentum) flux which are less affected by the boundary conditions. Here the Prandtl number ($Pr=0.63$) used is calculated from a quasi-linear flux-tube simulation at mid-radius using nominal parameters which is consistent with previous calculations \cite{strintzi08}.  

At the small values of $\rho_{*}$ seen in Ohmic plasmas, the value of the heat flux is known to tend towards the value calculated using a local model \cite{mcmil10}.  The flux-tube calculated value of the heat flux and its standard deviation at mid-radius are also plotted in the bottom-left (blue crosses and error bars) panel of Fig.~\ref{flows} showing agreement with the heat flux from the global calculation.  At the values of $\rho_{*}=0.0015$ used here the global simulation result tends toward the local flux tube result (blue crosses) \cite{mcmil10,candy04}, which is formally $\rho_{*}=0$.  This is seen here with the values of the two different simulations being within the error bars.  Also plotted is the momentum flux when an equilibrium $u'$ is imposed that is seen in the global simulation, in this case $u'=-0.8$  and we see that this is indeed of the same size as the residual stress seen in the global simulation.   The ion ($\chi_{i}=2.5$) and electron ($\chi_{e}=0.6$) heat diffusivities at mid radius agree reasonably well with those calculated via power balance, ($\chi_{i} \sim 2.4$ and $\chi_{e}\sim 0.2$  respectively) \cite{erofeev}.

\subsubsection*{$\rho_{*}$-scaling}

\begin{figure}
\begin{centering}
\includegraphics[width=6.3cm,clip]{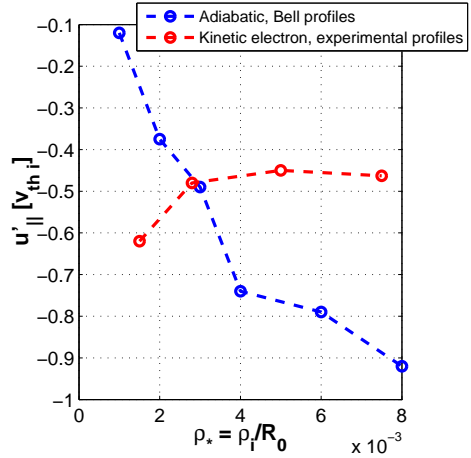}
\caption{The scaling of the mid-radius flow gradient, $u'$ as function of $\rho_{*}$ for simulations with kinetic electrons (red) and for adiabatic electrons (blue, taken from \cite{Buchholz14}). }
\label{rhscaling}
\end{centering}
\end{figure}

A series of global simulations were performed with the same profiles as used in the previous section, but with increasing values of the normalised gyro-radii to ascertain the scaling of the intrinsic flow.  The number of radial grid points and toroidal modes were scaled accordingly to maintain the same resolution.  Previous work showed that the flow gradient scaling, with adiabatic electron response, had two regions.  At small gyro-radius the gradient scales linearly and then saturates above $\rho_{*}=4\cdot10^{-3}$ \cite{Buchholz14}.  With kinetic electrons it is seen that the scaling of the flow gradients  with the normalised ion-gyro-radius is weak and that these two separate regions are not apparent.  The normalised flow gradient at mid radius ($u'$) as a function of $\rho_{*}$ is plotted Figure \ref{rhscaling}.  In fact an inverse scaling is seen at very small values of  $\rho_{*}$.  This could be attributed to the boundary conditions having less impact on the flow gradient in the centre of the computational domain as the simulation becomes more local in nature.

\begin{figure*}[ht]
\begin{centering}
\includegraphics[width=17.5cm,clip]{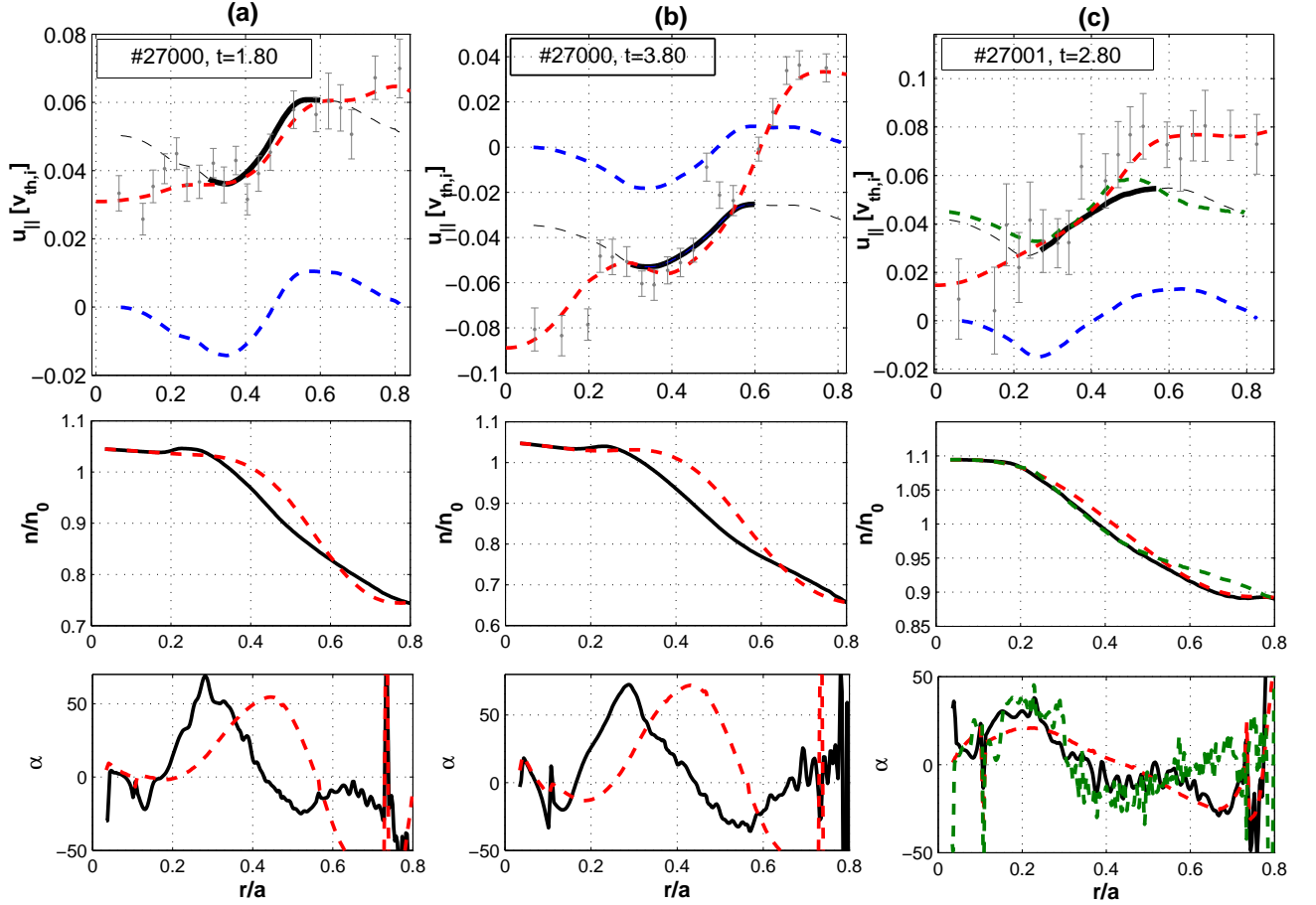}
\caption{(top) The parallel ion flow ($u_{||}$) for the final state of three global simulations for three time slices during discharge \#27000-27001 of AUG utilising $\rho_{*}=0.005$.  The top panel of (c) shows also the flow profile from an experimentally accurate, $\rho_{*}=0.0015$ simulation (green dashed).  The red dashed lines correspond to the CXRS generated experimental flow profiles with (in grey) the raw data and error bars.  The (blue) dashed lines represent the final state of a GKW turbulence simulation the (black) profiles represent the same profile but shifted to match the experimental value measured at the $r/R=0.1$.  The central region is highlighted as this is the region least influenced by the boundary conditions.  (Middle) The intial (red) and final (black) density profiles and (Bottom) the initial (red) and final (black) density curvature profiles.}
\label{flows2}
\end{centering}
\end{figure*}

In light of this weak scaling with $\rho_{*}$, a further series of simulations were performed for $\rho_{*}=0.005$ at 3 time slices (\#27000, $t=1.8s$ panel (a) in Fig.\ref{flows2}; \#27000, $t=3.8s$ (b); \#27001, $t=2.8s$ (c) ) within the discharge encapsulating two flow reversals to see whether the variation in flow gradient during the discharge is reproduced.  The final flow states, and the comparison with the experimental profile is plotted in Figure.~\ref{flows2} and the corresponding initial and final density profiles and curvature profiles are also plotted.  The comparison with the experiment is restricted to the middle of the simulation domain.  A reasonable agreement is seen in all three cases, however all three are hollow flow profiles which show varying degrees of profile hollowness.   

In these simulations it is observed that the flow profile evolves alongside an evolution of the density profiles.  In the middle and lower panels the corresponding density and density curvature profiles are plotted.  Here we define the profile curvature as the normalised second radial derivative of the density profile, 

\begin{equation}\alpha = -\frac{R_{0}^{2}}{n}\frac{\partial^{2}n}{\partial r^{2}}.
\end{equation} 
Both sets of plots show the equilibrium initial density profiles (in red) and the final state (in black).   The initial profiles are taken from experimental density profiles reconstructed with the integrated data analysis (IDA)  system.  The effect of density profile evolution will be discussed in depth in the following sections.

\section{Density evolution and its effect on profile shearing}
\label{d2n}

To maintain background temperature profiles, gradient driven global simulations utilise a Krook style operator of the form described in the Sec.~\ref{krook}.    In the simulations presented, a value of 
$\gamma_{k} = 0.028 v_{th i}/R_{0}$ is used for the rate coefficient.  This keeps the deviation in the temperature profile to less than 0.1\% from the equilibrium value.  This form of operator, however, does not act to maintain the density profile at its equilibrium form.  In simulations with adiabatic electrons this means that the density and temperature profiles stay very close (within 1\%) of the equilibrium profile, since radial particle transport is suppressed.  However, when electrons are treated kinetically, particle transport is finite, and thus the density profile is allowed to evolve (examples are shown in the top right panel of Fig.~\ref{d2ndx2fig}).   To reach a steady flow state the code is run on a transport time-scale, and as such,  the density profile evolves and relaxes.  
This relaxation occurs until, like the momentum flux, the radial particle flux at each radial location is zero.  The particle flux may also be written as a sum of a convective and diffusive component $R\Gamma_{i}/n = D_{i}R/L_{n} + RV_{i}$ \cite{angioni12,Mikk15}.  This relaxation of the density profile combined with the radial boundary conditions, which pin the density profile to the equilibrium value in the buffer regions, can introduce larger profile curvature while maintaining the same mean gradient, $R/L_{n}$.  

One of the dominant residual stress mechanisms is profile shearing.  This is a symmetry breaking mechanism whereby a poloidal tilting of the turbulent global mode structure is caused by the radial variation of equilibrium quantities.  The residual stress due to profile shearing, $\Pi_{res}=\Pi_{res}(n'_{e},n''_{e},T'_{i},T''_{i}...)$  \cite{Buchholz14}.  The residual stress evolves along with the density profile.   Ref.~\cite{Buchholz14} showed the effect of curvature on profile shearing using an adiabatic electron model over a variety of profile shapes, however the amplitude of the curvature was estimated as $(R/L_{ne})^2$. This can significantly underestimate its magnitude and its variation.   An increase in the local flow gradient could also be attributed to a reduction in the Prandtl number at mid-radius due to the variation in the equilibrium parameters.  However, flux-tube simulations  where the logarithmic density gradient is the only parameter varied between simulations ($\epsilon=0.12$, $R/L_{n} =  1.55 \rightarrow 0.5$) showed a negligible variation of the Prandtl number and as such this possible explanation can be excluded. 

\subsection{Simulation of discharge number \#32842}

A further simulation was performed of a more recent discharge, \#32842 at time, $t=2.16s$ ($I_{p}=0.83$ MA, $B_{T} = 2.39$T) in an attempt to see whether a peaked or flat flow profile was reproduced in a global simulation.  Experimental data taken from \#32842 is superior to that of \#27000 due to an improved CXRS diagnostic system as well as improved electron density measurements.  

\begin{figure}[ht]
\begin{center}
\includegraphics[width=8.5cm,clip]{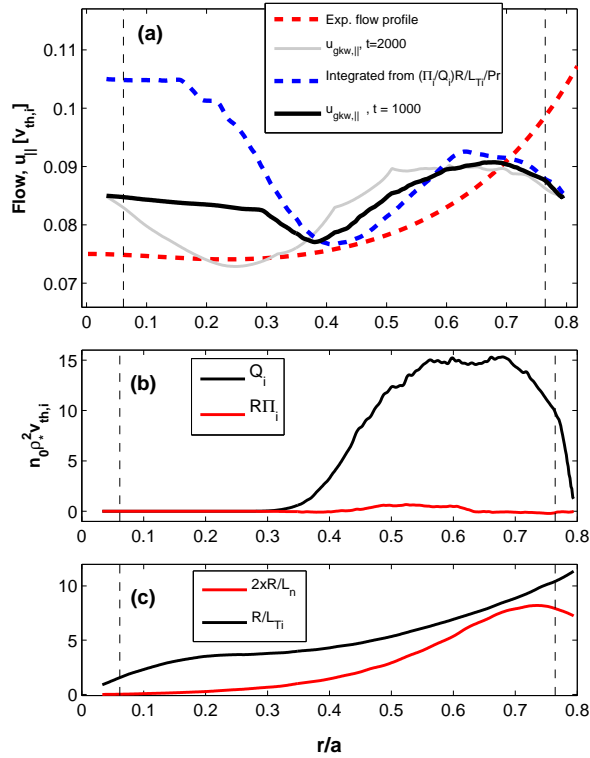}
\caption{(a)  The radial profiles of the turbulence generated flow profile at two different simulation time points (black, $t=1000R_{0}/v_{thi}$) and (grey, $t=2050R_{0}/v_{thi}$), as calculated from the residual stress (blue) and the experimentally measure profiles (red dashed). (b) The time averaged radial heat (black) and momentum (red) flux just after the linear overshoot of the simulation.  The vertical dashed lines denote the buffer regions where the turbulence is damped (c) The radial profiles of the logarithmic gradients in the equilibrium kinetic profiles.}
\label{leb1}
\end{center}
\end{figure}

The radial profiles of the flows, heat and momentum fluxes and kinetic gradients are shown in Figure~\ref{leb1}.  Apart from the kinetic and geometric profiles all the parameters are the same as in the previously described simulation.

In this simulation the curvature in the density profile is significantly smaller $\alpha \leq 24$ and has a different sign in the centre of the domain than in the previously simulated discharge and, as such, the residual stress (middle panel) is significantly smaller with respect to the heat flux.  It should also be noted that the curvature in the ion temperature profile is also significantly reduced ($\alpha\leq 30$ as opposed to $\alpha\leq 50$ in the previously described simulation) and that the residual stress component due to the radial variation of this is also smaller.  Correspondingly, the simulated flow gradient around $r/a=0.55$ is smaller ($u'=-0.3$) with respect to the flow gradient in \#27000.  This compares well with the reduction of the gradient seen in this time slice (red dashed line), however, it is also evident that flow profiles of this form are not well reproduced in the simulation even though the sign of the flow gradient is.  We note also that the experimentally measured flow amplitude is large ($u\sim0.08$) and that Coriolis pinch effects would be significant.  This would further reduce the amplitudes of the flow gradients, particularly in the outer region.  A previous study of residual stress mechanisms \cite{Hor17} showed that the contribution to $u'$ from the Coriolis pinch was approximately $u'\sim 0.01-0.06$ at $\rho_{T}=0.35$. This corresponds to a correction of around 0.1 in this simulated discharge  ($u=0.08$). 

Plotted are flow profiles at two different times in the simulation.  Firstly one at $t=1000 R_{0}/v_{thi}$ before the density profile has significantly relaxed, and a second profile later on, where the relaxation in the density profile produces an increase in flow gradient and reversal of sign at mid radius ($r/a=0.35$) due to the modification in the residual stress.  This will be discussed in more depth in the next section. 

\subsection{Sensitivity of residual stress component to density curvature.}

In this section the effect of density profile evolution on the residual stress is underlined by varying the equilibrium density profile used in the simulation described in Sec.~\ref{sec151} and by comparing the results from using both kinetic and adiabatic electron models.

\begin{figure}[ht]
\begin{centering}
\includegraphics[width=8.0cm,clip]{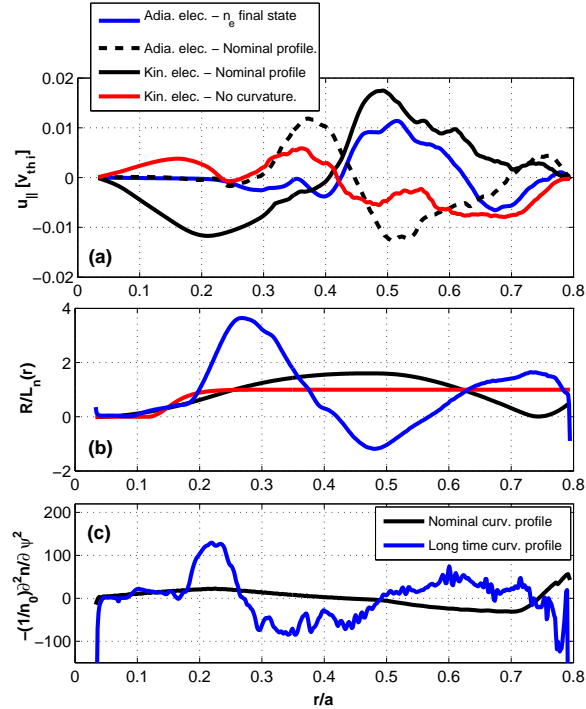}
\caption{(a) Flow profiles from a series of simulations with kinetic and adiabatic electron models, and variation of the density profiles. (black solid) Simulation with kinetic electrons and experimental profile, (black dashed) with adiabatic electron model, (red) kinetic electrons with a constant $R/L_{n}$ density profile and (blue) adiabatic electrons but with a density profile taken from a late-time relaxed state. (b) The profiles of $R/L_{n}$ used in the above simulations. (c) The profiles of $\alpha = -\frac{1}{n_{0}}\frac{\partial^{2} n_{e}}{\partial \psi^{2}}$ used in the above simulations. }
\label{d2ndx2profs}
\end{centering}
\end{figure}

A direct correlation between the magnitude of the perturbation of the density profile and the parallel flow gradients is seen in the simulations (panel (a) from  Fig.\ref{leb1} and Fig.\ref{d2ndx2profs}).  In both the simulated discharges as the density profile evolves, increasing its mean curvature, the local flow gradient increases accordingly.  This corresponds to an increased residual stress due to an increase in profile shearing.   If one calculates the curvature of the perturbed density profile ($-\frac{1}{n_{0}}\frac{\partial^{2} \tilde{n}}{\partial r^{2}}$) and equilibrium ($-\frac{1}{n_{0}}\frac{\partial^{2} n_{0}}{\partial r^{2}}$) separately (see panel (c) of Fig.~\ref{d2ndx2profs}), one sees that at late time the overall curvature is totally dominated by the perturbed component.  It should be also noted that the perturbation in the simulation at late time is much larger than those that would be within the experimental error bars of the measurements.  Moreover, the flows plotted in Fig~\ref{flows} are calculated when the density profiles are still close to nominal ($4-5\%$ perturbation away from the equilibrium), and as the density perturbation increases the flow gradient goes away from the experimentally measured value.

When the full experimental profiles are used one obtains a reversal in flow gradient at mid-radius when comparing adiabatic electrons (dashed black line) with a simulation with kinetic electrons (black solid line).  This is consistent with a reversal in residual stress sign simply by changing the electron model.   The role of the density profile on the generation of residual stress is emphasised by taking the relaxed density profile which produces a zero radial particle flux state after a long time, kinetic electron simulation and then using this as the equilibrium background density profile with the adiabatic electron model.  The density profile used and the final flow state from the adiabatic simulations are shown in panel (b) of Fig.~\ref{d2ndx2profs} (blue solid line).  It is evident that the sign of the gradient reverses once more to a state which has a similar flow gradient to the kinetic electron case with experimental density profiles (panel (c), black solid line).  The two different density profiles vary the residual stress enough to obtain a flow gradient reversal.

Finally, as a numerical experiment, a simulation with a constant density gradient and zero second derivative was performed while keeping all other profiles the same, particularly the ion temperature gradient, which is the predominant ITG mode drive.  In this case the component of the profile shearing residual stress due to the curvature is removed and this dramatically changes the flow profile seen, being enough to get a flatter and peaked profile as opposed to a hollow case as we normally see with kinetic electrons, signifying that the curvature, once more, is a dominant contributor to the residual stress. 

The density profile evolution also goes to explain why hollow profiles are always seen eventually in kinetic electron simulations, even when peaked or flat flow profiles are expected since the curvature of the profile due to its relaxation is always seen to evolve in the same direction.   While this makes direct comparison of the flow profile with the experiment more delicate, restricting analysis to a point in the simulation where the density perturbation is small, it does allow a measurement of the sensitivity of the flow gradient to small variations in the density profile.  Then it is possible to see if the variations in the density profile alone are large enough and flexible enough to explain the flow reversals seen in the experiment.

\begin{figure*}
\begin{centering}
\includegraphics[width=13.0cm,clip]{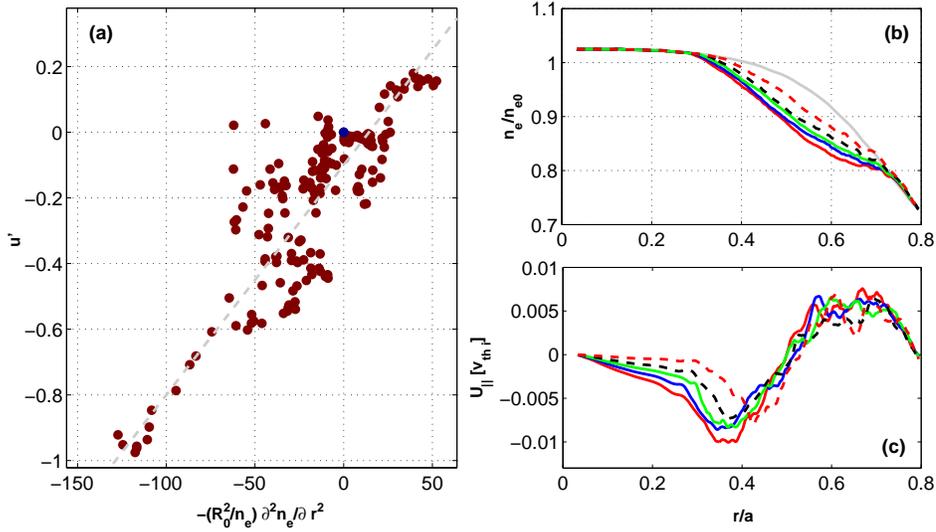}
\caption{(a) The flow gradient, $u'$ as function of the second derivative of the time averaged and spatially averaged density profile for multiple global turbulence simulations with electrons treated kinetically.  Snapshots of the density profile (b) and the corresponding flow profile (c)  }
\label{d2ndx2fig}
\end{centering}
\end{figure*}

The sensitivity is investigated by slicing the domain of all the simulations with experimentally relevant $\rho_{*}$ as previously described, into 4 different radial sections.   The time evolution of the flow gradient and the density profile during the simulation is then also divided into time windows.  Using this to sample the data, by averaging over the spatial and time windows, the flow gradient is plotted as a function of normalised second derivative of the density ($-R^{2}_{0}/n_{e}\frac{\partial^{2}n_{e}}{\partial r^{2}}$).  
The areas closest to the boundaries were also neglected from this analysis to minimise boundary effects.  The local flow gradient, $u'$ against the mean curvature  is plotted for simulations with kinetic electrons based around the above simulations in panel (a) of Fig.~\ref{d2ndx2fig}.    From a linear fit to this data we can estimate that, due to density profile curvature, and via profile shearing the residual stress can give
a flow gradient of,
\begin{equation}
u' \sim -0.008 \frac{R^{2}_{0}}{n_{e}}\frac{\partial^{2}n_{e}}{\partial r^{2}} = 0.008 \alpha.
\label{scalup}
\end{equation}
There is a large amount of scatter in the data as it also encompasses variations in the ion and electron temperature profiles and the safety factor  which are also influencers on the residual stress and their variation will also be captured in this analysis.  The relative strengths of the density and temperature gradients and second derivatives was investigated in Ref.~\cite{Buchholz14} and were found to be equal in magnitude, however opposite in sign.  Our result with kinetic electrons is consistent in its sign and amplitude with this result, whose prediction for the variation with density curvature would give a coefficient of approximately $u'=0.012\alpha$.  It should be noted that due to the boundary conditions, while the local gradient and curvature can vary significantly the mean density gradient across the domain is constant as the profile is pinned at the boundaries.  

The density relaxation process does vary significantly in amplitude with $\rho_*$.  The fractional perturbation of the density profile ($\delta n_{s}/n_{0}$) is inversely proportional to $\rho^{2}_*$ (if we assume a gyro-Bohm scaling of the transport coefficients,    $D_{i}\sim \rho_{*}$ \cite{Manf97}).  As we increase $\rho_*$, global effects reduce the particle flux in much the same way as the heat flux \cite{mcmil10}.  The ratio of $\Gamma_{i}/Q_{i}$ does not vary significantly.

The net effect is that the perturbation to the density profile required to reach a zero particle flux state is smaller at higher $\rho_{*}$.  Therefore, a smaller perturbation to the density profile is needed and as such the variation of residual stress and its corresponding change in $u'$ are smaller.

\subsection{Components of $k_{||}$}

Recent theoretical work has shown how global mode structures can be described in the local approximation \cite{XLu15}.  The global mode structure can be parameterised using the real and imaginary parts of the radial wavenumber, $k_r$, the parallel wavenumber, $k_{||}$ and the shift in the ballooning angle away from the outboard mid-plane $\theta_0$.  These are all components of a generalised ballooning mode description and can characterise the structure of a global turbulent mode in a local description.  These parameters are also intrinsically linked to the asymmetry in the mode and, therefore, also to intrinsic momentum transport.

\begin{figure}[ht]
\begin{centering}
\includegraphics[width=8.8cm,clip]{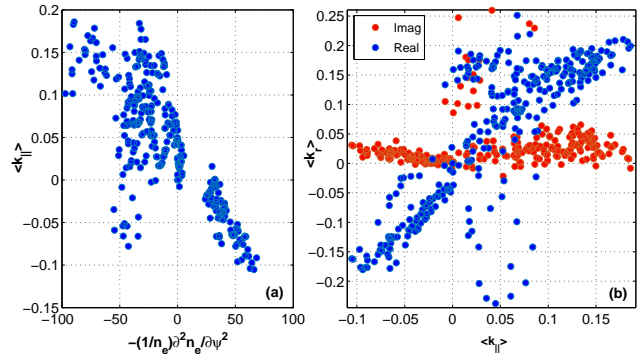}
\caption{(a) The variation of the spectrally averaged parallel wave-number ($<k_{||}R_{0}>$) against the density profile curvature. (b) The real (blue) and imaginary (red) components of the spectrally averaged radial wave number as a function of the parallel wave-number. }
\label{kparfig}
\end{centering}
\end{figure} 
These are defined and normalised in the following way in GKW,
\begin{eqnarray}
\langle \theta_{0} \rangle = \frac{1}{\langle |\phi_{n}|^{2} \rangle} \int \sum_{n} \theta|\phi_{n}|^{2} ds \\
\langle k_{r}^{R}R_{0} \rangle = \frac{1}{\langle |\phi_{n}|^{2} \rangle}  \Re \int  \sum_{n} \phi_{n}^{\dagger}\frac{\partial\phi_{n}}{d\psi} ds\\
\langle k_{r}^{I}R_{0} \rangle = \frac{1}{\langle |\phi_{n}|^{2} \rangle} \Im \int  \sum_{n} \phi_{n}^{\dagger}\frac{\partial\phi_{n}}{d\psi} ds\\
\langle k_{||}R_{0} \rangle = \frac{1}{\langle |\phi_{n}|^{2} \rangle} \int  \sum_{n} \phi_{n}^{\dagger}\frac{\partial\phi_{n}}{ds} ds
\end{eqnarray}
where the angled brackets denote a flux-surface averaged quantity and the sum is over the toroidal mode number.  Daggers denote a complex conjugate since the potential is decomposed into Fourier modes in the toroidal direction.  The real and imaginary parts of $k_{r}$ can be linked to specific residual stress mechanisms \cite{XLu15},  the real part being related to profile shearing effects, while the imaginary part is determined by turbulence intensity gradient effects.  Therefore with a similar analysis to the one with the density profile, we are able to plot the variation of these two quantities with the density curvature.  This is shown in Fig.~\ref{kparfig}.  The left hand panel (a) shows how the parallel wave-number varies linearly with the density curvature parameter, while the right hand panel (b) shows that this, in turn, is directly proportional to the real part of the radial wave-vector and the imaginary part is a small, but constant component.  This disparity becomes larger as the curvature increases in the simulation, thereby increasing the real component of $k_{r}$ while the imaginary part stays almost constant reflecting the fact that $k_{r}$ describes the strength of profile shearing and thereby increases with $\alpha$.  Therefore, it is possible to say that, for these simulations of Ohmic L-mode plasmas, profile shearing largely determines the intrinsic rotation profile.

\subsection{Application to experimental data}
\label{apptoexp}

The simulation results provide clear indications of a link
between the second radial derivative of the density profile
and the radial gradient of the intrinsic rotation angular velocity.  Estimates of second derivatives require diagnostics delivering accurate measurements at extremely high radial resolution. For the density profiles, these can be obtained by reflectometry diagnostics, and in particular by the ultra fast swept reflectometer (UFSR) \cite{clairet}, which became recently available on ASDEX Upgrade for a limited time \cite{medved}.  

This motivates the study of measurements of the density profiles with high time and radial resolutions in recent Ohmic heated discharges featuring density ramps across the LOC-SOC transition \cite{lebschy}. In these plasma conditions, measurements typically have a radial resolution of between 5 and 10mm in the core, and a time resolution of 0.5ms.

\begin{figure}[ht]
\begin{centering}
\includegraphics[width=7.0cm,clip]{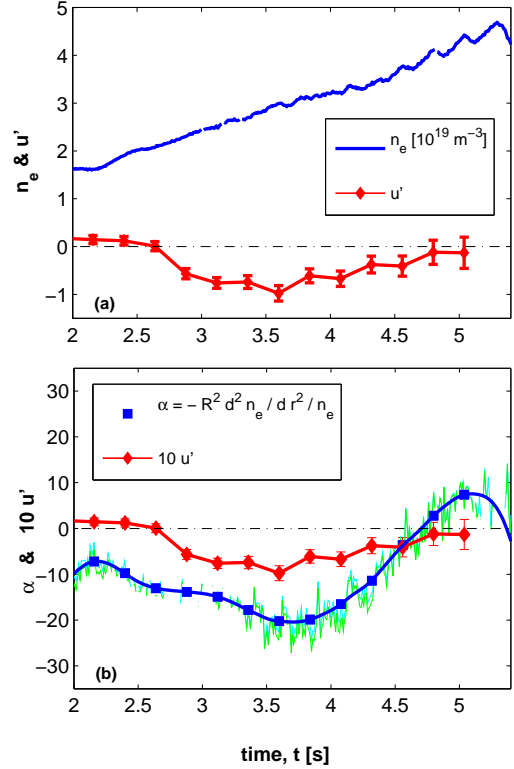}
\caption{Time evoluton of the local density and of the normalized toroidal angular velocity gradient $u'$ as a function of time (top) and time evolution of the normalized second derivative $\alpha$ (bottom), during the current flat top phase of the AUG discharge \#32842, at 800 kA and 2.4 T.  All quantities have been computed as radial averages in the interval $0.4 < \rho_{tor} < 0.6$.  The time evolution of $u'$ is also plotted in the bottom panel (with a factor of 10 for plotting purposes).}
\label{expscalfig}
\end{centering}
\end{figure} 

\begin{figure}[ht]
\begin{centering}
\includegraphics[width=8.0cm,clip]{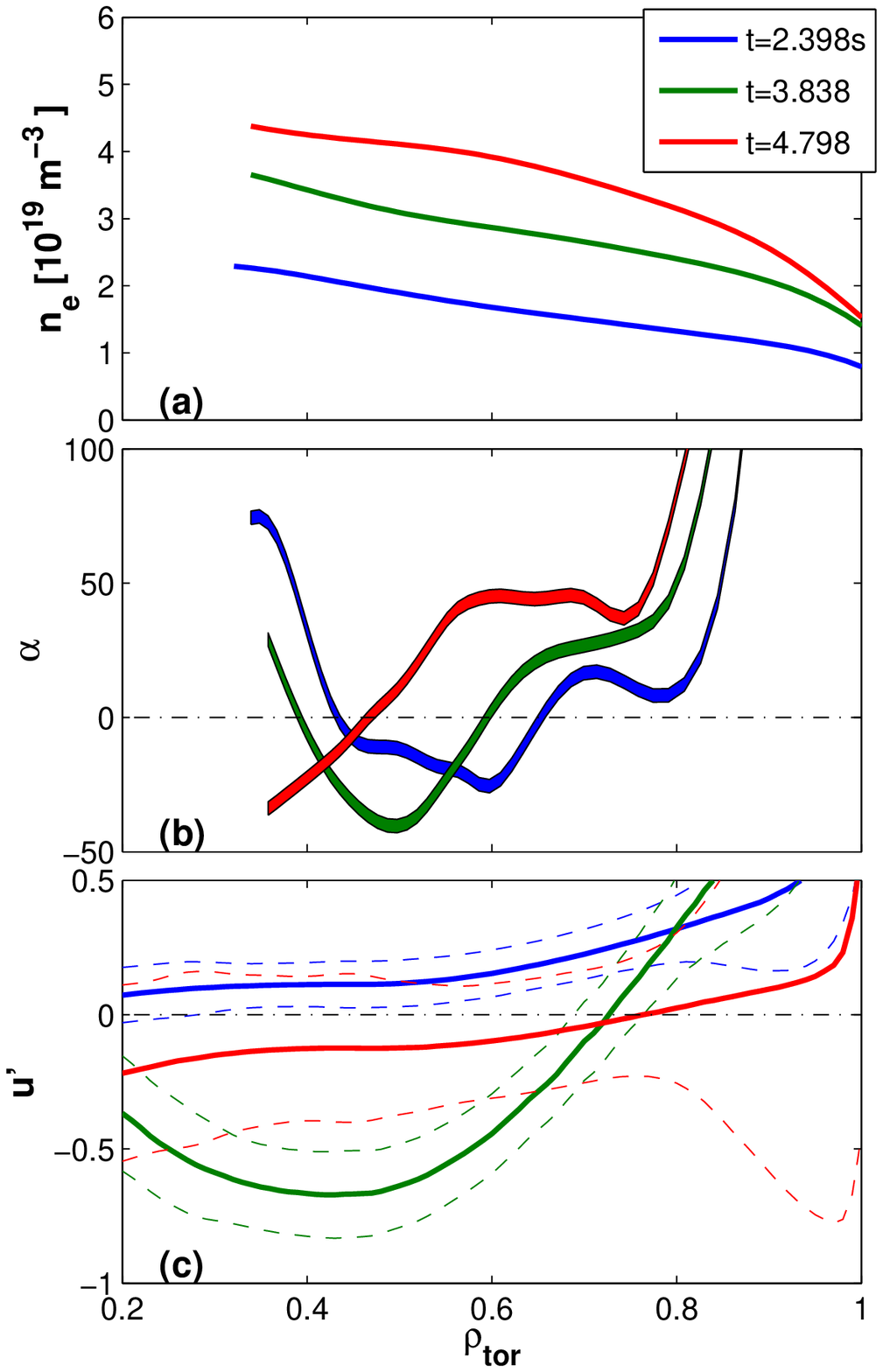}
\caption{Profiles of the density (a), the normalized second derivative of the 
density $\alpha$ (b), where the error bar in the profiles of the alpha parameters, as provided by 
the UFSR, are based on the sensitivity of the measurement of the time of flight of 
the probing wave to small variations of the measured density profile, and the normalized toroidal angular velocity gradient $u'$ (c) at three 
representative time slices of the AUG discharge \#32842, at $800 kA$ and $2.4T$.}
\label{expscalfig2}
\end{centering}
\end{figure} 

With this data, the second derivatives of the density profiles have been computed by considering time averaged profiles with a time resolution of 2.5 ms (averaging over sets of 5 profiles). The radial derivatives are computed with respect to the geometrical minor radius defined as half width of the mid-plane diameter.  The precision of the measurement of the second derivative can be evaluated by computing the sensitivity of the time of flight of the probing wave to small variations of the measured density profile.  The precision of the time of flight measurement is about $\pm 0.1ns$, which implies that variations in the profile curvature parameter, $\alpha$ of the order of $\Delta\alpha = 5$ can be resolved by the diagnostic.

In Fig. \ref{expscalfig} the time evolutions of the local density and of the radial average of the second derivative in the interval $0.4 < \rho_{tor} < 0.6$ are compared to that of the normalised radial gradient of the intrinsic rotation angular velocity $u'$, in the current flat top phase of the 800 $kA$ AUG discharge $\#$32842 at $2.4T$.  The second derivatives of individual density profiles, defined as $\alpha=-(d^2n_e/d r^2)R^2/n_e$, can exhibit large oscillations both along the minor radius, with values locally exceeding $\pm 50$, and in time, with also comparably large oscillations (as shown in Fig. 11 of Ref [21]).

However, when time averages over multiple profiles are considered, more regular behaviour is obtained, which allow us to compare the time evolution of the second derivative of the density profile with that of the first derivative of the toroidal rotation.  In the top panel of Fig.~\ref{expscalfig}, we observe that, with increasing density, the measured values of $u'$ vary from approximately zero at low density to almost -1 at intermediate densities to eventually increase at higher density back to values close to zero. This non-monotonic behaviour of the intrinsic rotation profile shape \cite{ang11,RMac14} is similar to that of the second derivative of the density profile, plotted in the bottom panel of Fig. \ref{expscalfig}, which also exhibits a non-monotonic behaviour with increasing density.  This observation supports the theoretical result that the curvature of the density profile is one of the parameters which play a role in determining the shape of the intrinsic rotation profile.  In Fig.~\ref{expscalfig2}, three profiles of the density, of the normalized second derivative $\alpha$, as well as of the corresponding normalized radial gradient of the toroidal rotation $u'$ are presented, at three representative time slices of the time evolution of the discharge \#32842 as reported in the legend.
We observe that the time slice at $3.838s$ is the only one developing a significant second derivative of the density profile, and correspondingly exhibits a negative value of the normalized toroidal angular velocity gradient parameter $u'$.  The minimum value of the $\alpha$ parameter, reaching -40, is consistent with the values obtained in the numerical simulations and which correspondingly develop similar local values of $u'$, below -0.5, as shown in Fig.~\ref{flows2} and Fig.~\ref{d2ndx2profs}.

\begin{figure}[ht]
\begin{centering}
\includegraphics[width=8.5cm,clip]{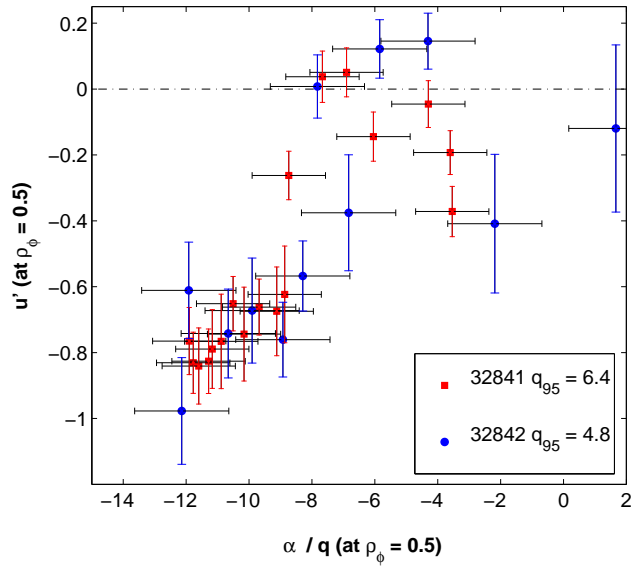}
\caption{Scatter plot of the normalized toroidal angular velocity gradient $u'$ as
a function of the  normalized second derivative of the density $\alpha$ 
divided by the local safety factor. All quantities have been computed as radial 
averages in the interval $0.4 < \rho_{tor} < 0.6$.}
\label{expscalfig3}
\end{centering}
\end{figure}

This observation is also consistent with the results presented in Fig.~\ref{d2ndx2fig}.  Fig.~\ref{expscalfig3} shows the scatter plot of $u'$ as a function of $\alpha$, evaluated in the same radial window $0.4 < \rho_{tor} < 0.6$, for two discharges, the \#32842 ($800 kA$) and the \#32841 ($600 kA$), where UFSR density profiles are available.  The local values of the $\alpha$ parameter have been divided by the local value of the safety factor, with the idea that it is the local value of the parallel wave number that mainly determines the residual stress. This normalization allows the scatter plots from the two separate discharges two overlap.  A clear trend is observed, where the largest negative values of $u'$ are observed only in correspondence to the largest negative values of $\alpha$.  We also observe that a large scatter is obtained for $\alpha/q$ values larger than -10, suggesting that the second derivative of the density profile should not be considered the only key parameter, as also indicated by Fig.~\ref{expscalfig} in the time window before 3s.  Also the corresponding plot obtained from the simulation results and presented in Fig.~\ref{d2ndx2fig} shows a significant level of scatter for values of $\alpha > -50$, which suggests that also in the simulations other parameters are important.

Finally, there are important limitations which should be mentioned and which might limit the extent to which quantitative comparisons can be performed.  We reiterate that in the experiment, it is the boron rotation profile that is measured, but here modelled is the main ion rotation, however the difference is known to be small, particularly in the intrinsic flow gradient.  Similarly, the electron density profile is measured and we assume that the ion density profile is identical.  However this neglects the boron density.  Boron concentration profiles in AUG are measured to be approximately flat, thereby one can assume that the profile features in the electron density profile, $n_{e}$ are also present in the main ion density profile, $n_{i}$.  This implies that one can not go much beyond a qualitative comparison with experimental data especially as accurate boron density measurements are not available for these discharges.

\section{Conclusions}
\label{concs}

Various mechanisms of toroidal momentum transport in the ASDEX Upgrade tokamak were studied using gyro-kinetic turbulence simulations.  It was shown in \cite{Hor17} that the combined effects of neoclassical 
flows, $E\times B$ velocity flow shear, Coriolis pinch, up-down asymmetry in the equilibrium and higher order corrections to the parallel derivatives are insufficient to explain the flow gradients that are measured.

In this work, global simulations of a selected subset of the discharge database show a good reproduction of the flow gradient sign and magnitude measured in AUG experiments.  Symmetry breaking mechanisms included in these simulations include profile shearing and turbulence profile effects.   The gradients achieved from these mechanisms are enough to make up the short fall seen in the previous flux-tube analysis.   However, it is seen that with kinetic electrons, simulations develop to give hollow profiles even when the experiment shows a flat profile.   Simulation of a flat flow case does show a marked reduction of intrinsic flow gradient due to a reduced residual stress, however, the qualitative agreement in this case is poor. 

Time evolution of the density profile due to finite particle fluxes in simulations go towards explaining this.  Density profile relaxation forces the simulation to give a hollow flow profile,  but also enables a quantification of the strength of curvature in the density profile on profile shearing as a residual stress mechanism.  Furthermore, it outlines the complex interaction between different transport channels, in this case the particle flux and momentum flux via the density profile and goes towards explaining the correlation seen between $R/L_{n}$ and flow reversals in some experiments \cite{ang11,Camen16}.  Using this time evolution of the density profile, it is shown that variation of curvature can strongly vary the residual stress and that variations in the density profile which are consistent in magnitude with experimental measurements, can explain the variation in the observed values of flow profiles measured on AUG.  While the second derivative of the density profiles clearly plays an important role in the development of hollow intrinsic rotation profiles, it is also clear that it can not be considered the only key parameter, as different degrees of hollowness of the rotation profile can be predicted without large variations in the value of the density profile curvature.

Highly accurate experimental measurements of the density profiles recently obtained from the ultra-fast swept reflectometer show elements of consistency with the results provided by numerical simulations, both in terms of the trend of the radial gradient of the intrinsic toroidal angular velocity as a function of the second derivative of the density profile and of the required values of these second derivatives.  At the same time, experimental results also show that the curvature of the density profile should not be considered  to be the only key parameter determining intrinsic flow profile hollowness.  Thereby, this theoretical work strongly motivates the investigation of means by which second derivatives of all the kinetic profiles can be measured, as these might have consistent behaviour with respect to the temporal evolution of plasma discharges that can be  directly linked to  highly reproducible, intrinsic rotation behaviour ubiquitously observed in ohmic L-mode tokamak plasmas.

\ack
Very useful conversations with Timothy Stoltzfus-Dueck and Yann Camenen are gratefully acknowledged.  The authors would like to thank F. Clairet and C. Bottereau
for their assistance and support in the installation and operation
of the reflectometry diagnostics.

Many of the simulations were performed on the HYDRA computer at the Max Planck Computing \& Data Facility centre (MPCDF), Garching bei M\"{u}enchen.  Part of this work was carried out using the HELIOS supercomputer system at Computational Simulation Centre of International Fusion Energy Research Centre (IFERC-CSC), Aomori, Japan, under the Broader Approach collaboration between Euratom and Japan, implemented by Fusion for Energy and JAEA.  


\end{document}